\documentclass[10pt]{article}
\usepackage[utf8]{inputenc}
\usepackage[T1]{fontenc}
\usepackage{lmodern}
\usepackage{amsmath,amssymb,booktabs,microtype,graphicx,placeins,float}
\usepackage[margin=1in]{geometry}
\usepackage[round,authoryear]{natbib}
\usepackage[hidelinks]{hyperref}
\usepackage{orcidlink}
\newcommand{\KL}{\mathrm{KL}}
\newcommand{\WQ}{W_Q}
\newcommand{\WK}{W_K}
\newcommand{\WV}{W_V}

\title{Quality Recovery for Quantized KV Caches via Low-Rank Attention Adaptation}
\author{%
Seifeldin Abdellatif \orcidlink{0009-0004-8879-9335}\\
College of Engineering, Al Ain University\\
\texttt{contact@saifmb.com}
}

\begin{document}
\maketitle

\begin{abstract}
Low-bit key--value (KV) caches reduce the memory required for autoregressive
decoding, but the resulting quality loss depends on the model and quantizer.
We keep the quantizer fixed and distill the floating-cache model's behavior into
low-rank Q/K/V projection updates while the student executes a physically
packed incremental cache. Across three seeds, 4-bit affine-cache adapters recover
$54.24\%\pm2.47\%$ of the held-out perplexity gap on TinyLlama-1.1B and
$75.96\%\pm4.04\%$ on Gemma-4-12B. On the same frozen NF4
Llama-3.1-8B base, one validation-selected run per quantizer recovers $60.42\%$
under KIVI
K2V2 and $37.61\%$ under KVarN K4V2, while preserving 180-case associative
retrieval. Gemma's score on an official 4K/8K RULER subset rises from 42.80
with the unadapted 4-bit cache to 48.33 after adaptation (46.15 floating), with
substantial task heterogeneity. Finally, a 2-bit rank--token sweep reduces
TinyLlama's 2-bit PPL from 576.10 to
$11.4000\pm0.0059$ across three seeds, versus 10.3988 floating, but restores
only 11--12 of 180 retrieval cases. These results show that low-rank projection
adaptation can recover held-out quality across fixed cache formats, while
perplexity recovery need not restore long-context retrieval.
\end{abstract}

\section{Introduction}

Autoregressive inference stores keys and values for later attention, so its KV
cache grows with sequence length and batch size. Low-bit representations reduce
this memory and bandwidth cost, but axes, outliers, residual precision, and
kernels all affect the remaining quality gap \citep{kivi,kvquant}. We ask
whether a small post-training update can recover loss after engineering
constraints fix the format. Low-rank updates alter the queries reading the
cache and keys and values written to it. We distill the floating-cache model
into a student executing the deployed packing and reconstruction path.

The base weights and quantizer remain fixed, calibration uses unlabeled text,
and only attention adapters are optimized. Dense-base updates use the standard
LoRA merge; the reported NF4 runs retain explicit adapters, while
quantization-aware LoRA provides a route to a single merged low-bit model
\citep{lora,qalora}.
We separate cache storage from adapter quality: the cache representation
determines storage, whereas held-out evaluation measures adapter recovery. We
therefore report absolute perplexity, normalized gap recovery, and long-context
behavior.

We evaluate three model scales, 2/4-bit affine and symmetric caches, and two
established quantizers: KIVI and KVarN. Primary affine comparisons use three
training seeds; placement and established-quantizer studies use one. A 2-bit
rank--token grid probes capacity, while NIAH and RULER test whether perplexity
recovery transfers to long-context behavior. Across the tested formats,
projection adaptation improves held-out quality, although attainable gains and
calibration requirements vary by operating point.

\section{Related Work}

\paragraph{Cache quantization.}
KIVI quantizes keys per channel and values per token while retaining a recent
high-precision residual window \citep{kivi}. KVQuant adds pre-RoPE key coding,
nonuniform datatypes, and sparse outliers \citep{kvquant}; SKVQ combines channel
reordering, clipping, and a recent window \citep{skvq}; and KVTuner searches
layer-specific K/V precision \citep{kvtuner}. Rotation-based methods reshape
the error before coding: RotateKV protects sinks and uses calibrated rotations
\citep{rotatekv}, whereas KVarN applies a fixed Hadamard transform and
dual-axis variance normalization \citep{kvarn}. These methods design the
representation. We instead optimize the model after the representation is fixed,
and experimentally cover grouped affine/symmetric coding, KIVI, and KVarN.
BitDecoding demonstrates that KIVI-style low-bit caches can be paired with
specialized tensor-core decoding kernels \citep{bitdecoding}. Our experiments
use KIVI's format with framework attention rather than specialized kernels.

\paragraph{Cache-error correction.}
GEAR augments a quantized cache with low-rank and sparse residuals
\citep{gear}; STAR-KV combines learned low-rank cache projections with mixed
precision \citep{starkv}. KVLinC integrates a rotated 2-bit representation,
trained feature maps, and recurrent correction state inside attention
\citep{kvlinc}. These methods allocate representation or operator state to
correction. Our post-hoc procedure leaves the cache format and attention
operator unchanged: ordinary Q/K/V projection updates absorb part of the error
and can be folded when the base projection is dense.

\paragraph{Low-rank adaptation and evaluation.}
LoRA supplies the foldable low-rank parameterization \citep{lora}, and QLoRA
shows that adapters can be optimized over frozen 4-bit NormalFloat weights
\citep{qlora}. We use these mechanisms for self-distillation rather than task
tuning. Because simple needle retrieval can conceal broader long-context
failures, we complement NIAH with RULER tasks spanning retrieval, variable
tracking, and aggregation \citep{ruler}.

\section{Method}

For one attention layer and hidden states $H$, write
\begin{equation}
 Q=H\WQ^\top,\quad K=H\WK^\top,\quad V=H\WV^\top .
\end{equation}
For a selected projection $W\in\{\WQ,\WK,\WV\}$, we train
\begin{equation}
 W'=W+\frac{\alpha}{r}BA,
 \label{eq:lora}
\end{equation}
where $A$ and $B$ have inner dimension $r$ and pretrained $W$ is frozen. We
evaluate Q-only, K-only, QK, and QKV placements.

\subsection{Physical incremental cache}

The student receives a cache whose persistent payload consists of bit-packed
codes plus scale metadata. For tensors of shape $[B,H,T,D]$, the affine variant
partitions the final head dimension into groups of 64. For group $g$,
\begin{equation}
 \begin{aligned}
 s_g&=\max\!\left\{\frac{\max(x_g)-\min(x_g)}{2^b-1},10^{-8}\right\},\\
 c_g&=\operatorname{round}\!\left((x_g-\min(x_g))/s_g\right).
 \end{aligned}
 \label{eq:affine}
\end{equation}
We clip $c_g$ to $[0,2^b-1]$ and store packed unsigned codes, an FP16 scale,
and an FP16 offset. The symmetric control stores a maximum-magnitude scale and
no offset. Both keys and values are quantized unless stated otherwise.

At inference, new states are packed and appended; the persistent cache retains
no unquantized copy. During calibration, autograd retains source activations for
the STE, but these are not part of the deployed cache state. Evaluation consumes
128-token chunks, exercising cache updates and reads. Reconstruction feeds
framework attention rather than a fused low-bit kernel.

Two established formats complete the study. KIVI K2V2 uses asymmetric
per-channel keys, per-token values, group size 32, and a 128-token FP16
residual. KVarN K4V2 uses group size 128, protects the first 128 sink tokens,
and applies a fixed normalized Hadamard transform followed by eight iterations
of dual-axis variance normalization. Validation against fixed official
revisions exactly matches KIVI's reconstructed tensors and KVarN's packed
records. In all cases, the deployed cache footprint comprises packed codes,
scale or offset metadata, and any format-specific FP16 residual or sink state.

\subsection{Quantizer interface and compatibility}

Let $\mathcal{R}_{\phi}$ denote the complete fixed cache path: coding,
metadata, residual or transformed state, and the reconstruction presented to
attention. Calibration replaces $(K,V)$ by $\mathcal{R}_{\phi}(K,V)$ in the
student forward pass.

\paragraph{Compatibility observation.}
If deployed attention with fixed $\mathcal{R}_{\phi}$ runs during calibration,
Equation~\ref{eq:kl} is well-defined. An identity straight-through estimator
(STE) carries gradients through hard coding to Q/K/V adapters. Folding
Equation~\ref{eq:lora} changes the projection feeding $\mathcal{R}_{\phi}$, not
the cache format.

The same interface can accommodate additive dense--sparse schemes such as
KVQuant or GEAR when their full reconstruction runs during calibration. A
stateful method such as KVLinC must instead be treated as a complete attention
operator. The experiments below cover affine and symmetric coding, KIVI, and
KVarN; recovery and optimization behavior remain empirical properties of each
format.

\subsection{Calibration objective}

For calibration sequence $x$, the teacher disables adapters and uses no
quantized cache. The student enables adapters and the packed cache. We minimize
\begin{equation}
 \mathcal{L}(x)=\KL\!\left(
 p_{\mathrm{float}}(\cdot\mid x)\;\Vert\;
 p_{\mathrm{packed}}(\cdot\mid x;A,B)\right).
 \label{eq:kl}
\end{equation}
Newly appended reconstructed states use a straight-through gradient:
$\hat{x}+x-\operatorname{stopgrad}(x)$. Forward attention therefore sees the
stored reconstruction while K and V adapters receive an identity backward path.
Q-only adapters do not require this path.

The objective does not invert quantization. It searches a restricted correction
space. For example, QK adaptation changes the unquantized logit product by
\begin{equation}
 \begin{aligned}
 &(Q+\Delta Q)(K+\Delta K)^\top-QK^\top\\
 &\quad=\Delta QK^\top+Q\Delta K^\top+\Delta Q\Delta K^\top .
 \end{aligned}
\end{equation}
The actual quantized map is piecewise and model dependent; this expansion only
explains why query and key changes can be complementary.

\subsection{Deployment overhead}

For a projection $W\in\mathbb{R}^{d_{\mathrm{out}}\times d_{\mathrm{in}}}$,
an explicit rank-$r$ adapter adds $r(d_{\mathrm{in}}+d_{\mathrm{out}})$
parameters and two low-rank matrix multiplications. At inference, this is the
same number of additional multiply--accumulates (MACs) per processed token.
These counts characterize static compute and storage, not wall-clock latency:
small auxiliary matrix multiplications can incur disproportionate launch and
memory-access costs depending on hardware, batch size, and kernel fusion.

The Gemma QK rank-32 adapter \citep{gemma4} contains 21.889M parameters, 0.18\% of the
nominal 12B model. Because the reported base uses NF4, its 87.6 MB of FP32
adapter weights equal 1.46\% of the nominal 6.0 GB raw 4-bit weight payload;
this comparison excludes NF4 metadata and runtime buffers. Explicit execution
adds 21.889M MACs per processed token.
This explicit branch accompanies $75.96\%\pm4.04\%$ held-out PPL-gap recovery.
The TinyLlama 2-bit rank-48 QKV adapter contains 9.191M parameters (0.84\% of
the nominal 1.1B model). Its dense-base update can be merged as
$W\leftarrow W+BA$, removing the adapter branch and leaving projection shapes
unchanged; this is the standard LoRA deployment with no adapter-side inference
latency \citep{lora}. The reported NF4 Gemma and Llama-3.1 runs retain explicit
adapters. QA-LoRA demonstrates that quantization-aware low-rank training can
instead integrate the base and adapter weights into one low-bit model
\citep{qalora}, providing a concrete merge-aware deployment path.

\section{Experimental Protocol}

\paragraph{Main models and training.}
TinyLlama-1.1B \citep{tinyllama} uses FP16 base weights, a rank-8 QK adapter
($1.126$M parameters), and WikiText-2 \citep{wikitext}. Each seed sees 256 sequences of 512
tokens for ten epochs, or 1.31M context tokens, at learning rate
$3\times10^{-4}$. After the placement sweep, we repeat the same recipe with
three rank-6, $\alpha=12$ QKV adapters ($1.149$M parameters). Gemma-4-12B
\citep{gemma4} uses frozen NF4 base weights, a rank-32 QK
adapter ($21.889$M parameters), and WikiText-103. Each seed sees 52,000
sequences of 384 tokens once (19.97M context tokens); the final 64 positions per
sequence are supervised, for 3.328M matched positions. Its learning rate is
$10^{-6}$. Gradient accumulation is eight. The models use different calibration
budgets, so cross-model differences are
interpreted separately rather than as a controlled scaling law. Both main
experiments use seeds $\{0,1,2\}$ and fixed within-model controls.

\paragraph{Established quantizers.}
On a fixed Llama-3.1-8B NF4 base \citep{llama3}, KIVI K2V2 and KVarN K4V2 each
use a rank-32, $\alpha=64$ QKV adapter. A 1,024-sequence pilot selected the learning rate
$10^{-5}$ from $\{10^{-6},3\!\times\!10^{-6},10^{-5}\}$ using WikiText-103
validation KL. Full runs expose the final 64 positions of 512-token sequences,
use seed 0, and save checkpoints after 1.25M, 5M, 10M, and 20M input tokens.
We select by validation PPL and evaluate the chosen checkpoint once on test.

\paragraph{Two-bit capacity study.}
For the TinyLlama 2-bit affine condition, we train QKV adapters at ranks
$\{6,12,24,48\}$ with $\alpha/r=2$. All ranks see the same shuffled
WikiText-103 calibration windows and are evaluated after 1.31M, 5.00M, 10.00M,
and 20.00M context tokens. We select one checkpoint per rank by full
WikiText-103 validation PPL, then evaluate the four selected checkpoints once
on the test split and on 180 shared NIAH cases. Rank selection uses a one-sided
exact sign test against the shared unadapted cases with Holm correction across
four ranks. The selected rank is retrained with two additional seeds and
evaluated on a fresh 180-case NIAH suite that was not used for selection.

\paragraph{Evaluation.}
We report incremental-cache perplexity on complete held-out windows: 340,326
scored tokens for TinyLlama on WikiText-2 and 291,463 for Gemma on
WikiText-103. Cache state resets at each 512-token TinyLlama window and
384-token Gemma window. Floating, unadapted packed, and adapted packed
conditions use identical tokens. Floating cache describes cache precision, not
necessarily base-weight precision. We summarize independently trained adapters
by their mean and sample standard
deviation. With three seeds, these statistics are descriptive.

\paragraph{Recovery metric.}
Let $P_f$, $P_q$, and $P_a$ denote floating-cache, unadapted packed-cache, and
adapted packed-cache perplexity. We report
\begin{equation}
 R=100\frac{P_q-P_a}{P_q-P_f}.
 \label{eq:recovery}
\end{equation}
This normalizes different gaps, but is unstable when $P_q-P_f$ is near zero and
can look favorable when every absolute PPL is poor. Tables retain absolute PPL.

\paragraph{Ablations.}
We test 2-, 4-, and 8-bit affine caches on TinyLlama; affine and symmetric
physical 4-bit caches; approximately parameter-matched projection placement on
TinyLlama and Llama-3.1-8B; and the rank--token grid above. Llama uses an NF4
base and each placement sees 1.31M training context tokens. Adapter counts
differ by at most 3.35\%. Placement comparisons use one seed, so their ordering
applies to these runs. The selected TinyLlama QKV conditions at 4 and 2 bits are
each repeated for three seeds.

\paragraph{Synthetic retrieval.}
Associative NIAH asks for a vault/code record among decoys at depths
$\{0.1,0.5,0.9\}$. Gemma uses 120 shared 1,024--8,192-token cases and a
demonstrated continuation prompt appropriate for its pretrained checkpoint.
TinyLlama uses its canonical chat template over 512--1,536 tokens. Bit-width
controls share 90 prompts; selected 4-bit QKV uses 180 fresh cases; and 2-bit
rank selection and three-seed confirmation use separate 180-case suites. A
case is correct only when the first generated eight-digit code equals the
target under greedy 24-token generation. Calibration contains no NIAH content.

We evaluate an official RULER subset \citep{ruler} on Gemma:
single-needle retrieval, multi-value retrieval, variable tracking, and common-word
extraction at 4K and 8K, with 25 examples per task--length cell (200 total).
We compare the floating cache, unadapted 4-bit affine cache, and seed-0 adapted
cache using the official string-match-all metric. This 200-example evaluation
covers four of RULER's 13 tasks.

\section{Results}

\subsection{Three-seed quality recovery}

Table~\ref{tab:main} gives the central result.
TinyLlama's packed cache raises PPL from 10.3988 to 10.7228; adaptation lowers it
to $10.5792\pm0.0017$, recovering $44.32\%\pm0.52\%$ of the induced gap.
The selected QKV placement lowers it further to $10.5471\pm0.0080$ and recovers
$54.24\%\pm2.47\%$ across three new training seeds.
Gemma moves from 26.5067 to 27.6208 and then to
$26.7745\pm0.0450$, a $75.96\%\pm4.04\%$ recovery. Every adapted PPL is
better than its corresponding unadapted control.

\begin{table}[H]
\centering
\caption{Main held-out PPL. Adapted values and $R$ are mean $\pm$ sample SD
over three seeds. Base denotes weight precision; caches are 4-bit affine.}
\label{tab:main}
\footnotesize
\renewcommand{\arraystretch}{1.08}
\begin{tabular*}{\columnwidth}{@{\extracolsep{\fill}}lrrrr@{}}
\toprule
Model / adapter (base) & Float & Packed & Adapted & $R$ (\%) \\
\midrule
Tiny, QK (FP16) & 10.3988 & 10.7228 & $10.5792\pm.0017$ & $44.32\pm.52$ \\
Tiny, QKV (FP16) & 10.3988 & 10.7228 & $10.5471\pm.0080$ & $54.24\pm2.47$ \\
\addlinespace[1pt]
Gemma, QK (NF4) & 26.5067 & 27.6208 & $26.7745\pm.0450$ & $75.96\pm4.04$ \\
\bottomrule
\end{tabular*}
\end{table}

\begin{figure}[H]
\centering
\includegraphics[width=0.98\columnwidth]{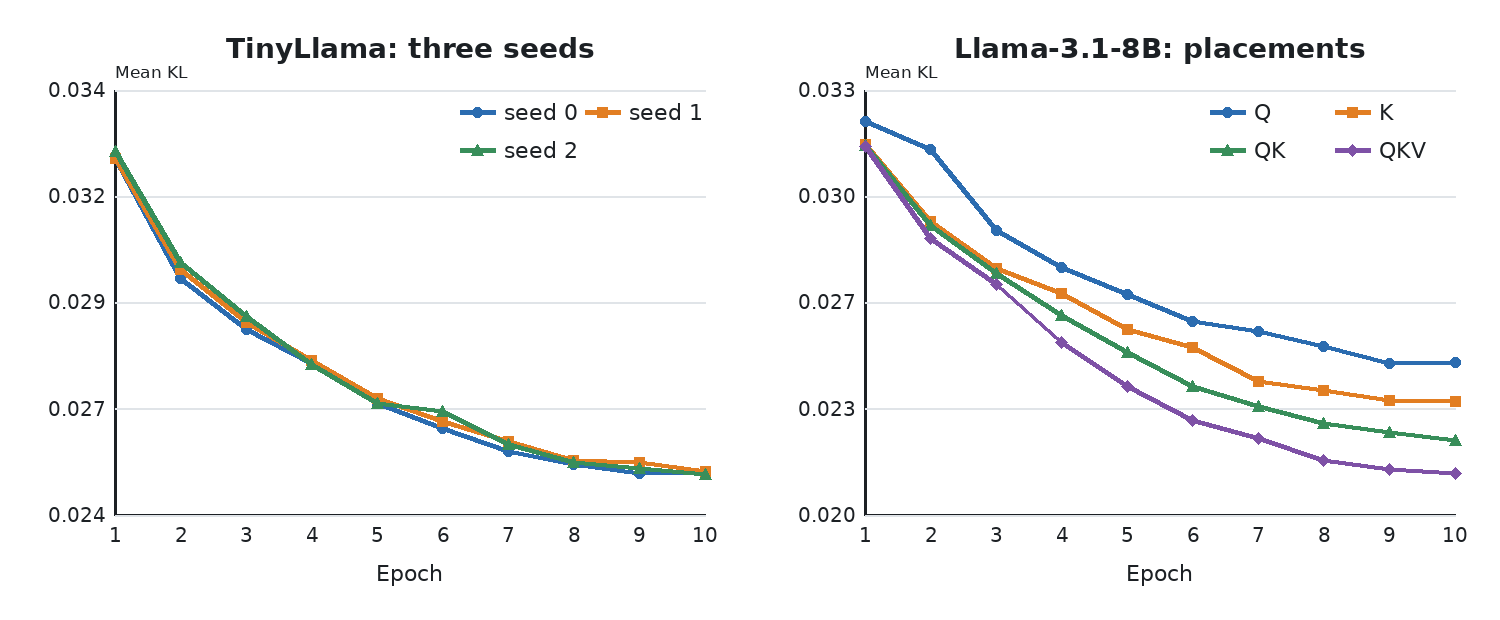}
\caption{Calibration loss decreases smoothly. TinyLlama's three QKV seeds
nearly overlap (left), while QKV finishes lowest among the Llama-3.1 placements
(right). Held-out quality is reported in Tables~\ref{tab:main} and
\ref{tab:placement}.}
\label{fig:convergence}
\end{figure}

Gemma attempted 6,500 optimizer updates per seed. Seeds 0--2 completed 6,342,
6,363, and 6,361 finite updates; 158, 137, and 139 non-finite updates were
skipped. Their final mean KL values (0.2152--0.2159) and held-out improvements
are close, but the skips are a numerical limitation. Figure~\ref{fig:convergence}
provides optimization traces: TinyLlama's three
QKV curves nearly coincide and flatten late. On Llama-3.1, every placement improves,
and QKV ends lowest. These curves describe optimization under the fixed budgets;
Tables~\ref{tab:main} and~\ref{tab:placement} report held-out quality.

\subsection{Where should the adapter be placed?}

Table~\ref{tab:placement} compares near-equal
budgets. Q and QK are nearly tied on TinyLlama, while K alone is weaker. Adding
V produces a larger improvement: QKV reaches 10.5383 PPL and 56.95\% recovery.
The Llama-3.1 replication has the same ordering at the top: QKV reaches 8.9034
PPL and 63.21\% recovery, compared with 44.01\% for QK.
Repeating TinyLlama QKV with two additional seeds gives recoveries of 56.95\%,
52.12\%, and 53.64\%, so its advantage over the QK mean is consistent across
three seeds. The Q, K, and QK alternatives were evaluated with one seed, and the
ablation is limited to TinyLlama and Llama-3.1, which share a broad decoder
family; Gemma was evaluated only with QK.

\begin{table}[!t]
\centering
\caption{QKV gives the strongest recovery under matched adapter budgets for
both models. Float and packed PPL are shared within each model; Params denotes
trainable adapter parameters.}
\label{tab:placement}
\footnotesize
\renewcommand{\arraystretch}{1.06}
\begin{tabular*}{\columnwidth}{@{\extracolsep{\fill}}lrrrr@{}}
\toprule
Place & Rank & Params (M) & Adapted PPL & $R$ (\%) \\
\midrule
\multicolumn{5}{@{}l}{\emph{TinyLlama-1.1B}: float 10.3988; packed 10.7228} \\
Q   & 12 & 1.081 & 10.5818 & 43.54 \\
K   & 22 & 1.115 & 10.6109 & 34.55 \\
QK  &  8 & 1.126 & 10.5787 & 44.47 \\
QKV &  6 & 1.149 & \textbf{10.5383} & \textbf{56.95} \\
\midrule
\multicolumn{5}{@{}l}{\emph{Llama-3.1-8B}: float 8.8195; packed 9.0475} \\
Q   & 11 & 2.884 & 8.9546 & 40.74 \\
K   & 18 & 2.949 & 8.9559 & 40.16 \\
QK  &  7 & 2.982 & 8.9472 & 44.01 \\
QKV &  5 & 2.949 & \textbf{8.9034} & \textbf{63.21} \\
\bottomrule
\end{tabular*}
\end{table}

\subsection{Recovery transfers to KIVI and KVarN}

\begin{figure}[H]
\centering
\includegraphics[width=0.98\columnwidth]{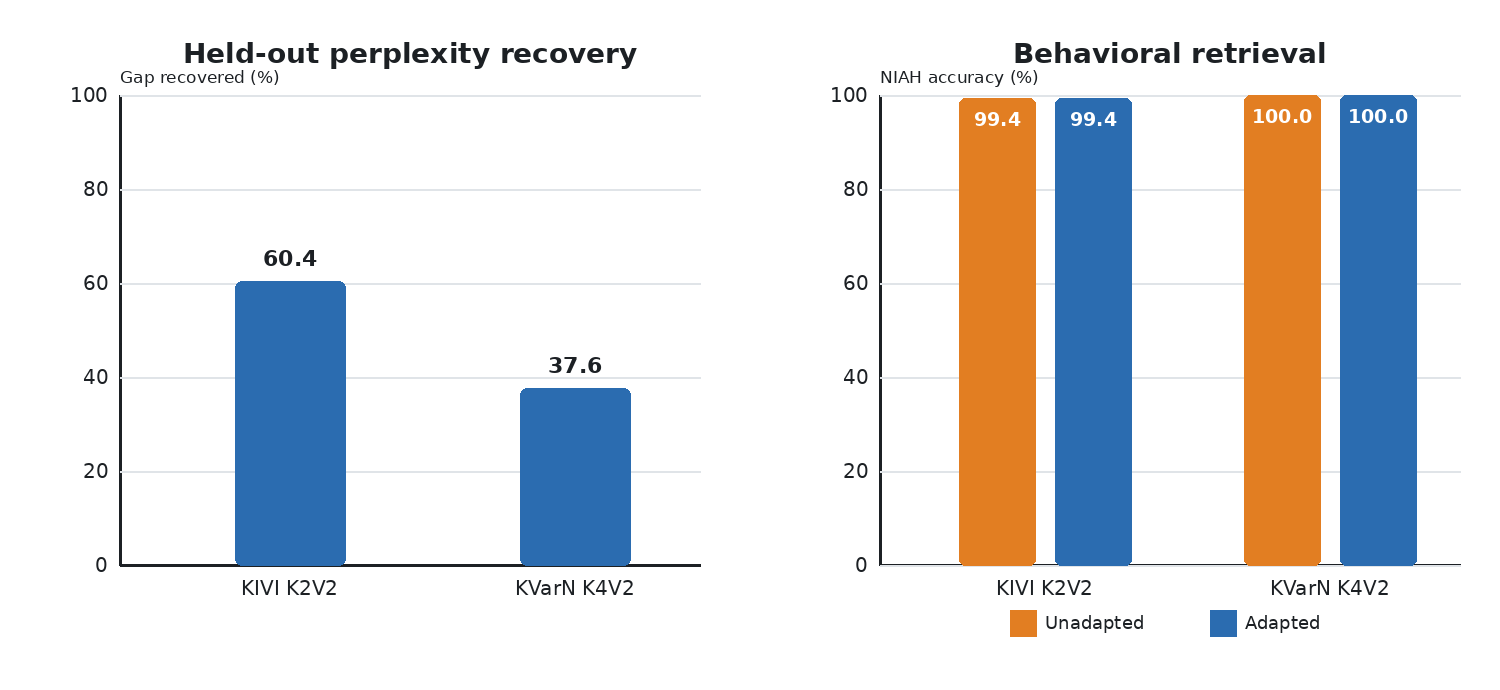}
\caption{Adaptation transfers to both established cache formats: it recovers
part of each held-out PPL gap (left) while preserving first-code NIAH accuracy
(right). Each method is one validation-selected training run.}
\label{fig:quantizers}
\end{figure}

Figure~\ref{fig:quantizers} holds the Llama-3.1-8B NF4 base, QKV rank, and
evaluation tokens fixed. KIVI K2V2 raises test PPL from 6.6637 to 7.1586;
the validation-selected 5M-token checkpoint lowers it to 6.8596, recovering
60.42\% of the gap. KVarN K4V2 has a smaller unadapted gap
(6.6637 to 6.7076); its selected 20M-token checkpoint reaches 6.6911, a
37.61\% recovery and a PPL reduction of 0.0165. Both adapters preserve the
unadapted NIAH score: 179/180 for
KIVI and 180/180 for KVarN.

Checkpoint trajectories differ. KIVI validation recovery at
1.25M, 5M, 10M, and 20M tokens is 63.90\%, 86.30\%, 83.04\%, and 84.24\%; its
optimum is reached early. KVarN improves throughout, from 97.43\% to 132.90\%
on validation, although its tiny validation denominator permits values above
100\%, and that super-recovery does not persist on test. Each quantizer uses one
validation-selected run, so these experiments demonstrate transfer to two
implemented pipelines but do not estimate variance or compare quantizer quality.

\begin{table}[H]
\centering
\caption{TinyLlama operating points ($P_f=10.3988$ throughout). Scaling the
2-bit adapter closes most of the PPL gap; the 8-bit recovery ratio is unstable
because its initial gap is near zero. Extended 2-bit values are mean $\pm$
sample SD over three seeds; other rows are one seed.}
\label{tab:operating}
\small
\renewcommand{\arraystretch}{1.05}
\begin{tabular*}{\columnwidth}{@{\extracolsep{\fill}}lrrr@{}}
\toprule
Setting & Packed & Adapted & $R$ (\%) \\
\midrule
Affine 2-bit, short QK & 576.1030 & 18.7758 & 98.52 \\
Affine 2-bit, 20M QKV & 576.1030 & $11.4000\pm.0059$ & $99.823\pm.001$ \\
Affine 4-bit & 10.7228 & 10.5787 & 44.47 \\
Affine 8-bit & 10.4012 & 10.3983 & $121.1^\dagger$ \\
Symmetric 4-bit & 11.2072 & 10.7570 & 55.69 \\
\bottomrule
\end{tabular*}
\par\smallskip
\footnotesize $^\dagger$Unstable normalization because the initial gap is
approximately zero.
\end{table}

\subsection{Bit width and adapter capacity}

Table~\ref{tab:operating} first shows why operating point matters. A short
rank-8 QK run at 2 bits lowers TinyLlama's PPL from 576.10 to 18.78, but remains
far above the 10.40 floating reference. At 4 bits, adaptation gives a modest
absolute and material relative improvement. At 8 bits, the initial gap is only
0.00235 PPL; the 121\% ratio reflects this near-zero denominator and is not an
informative measure of super-recovery.

\begin{figure}[!t]
\centering
\includegraphics[width=0.98\columnwidth]{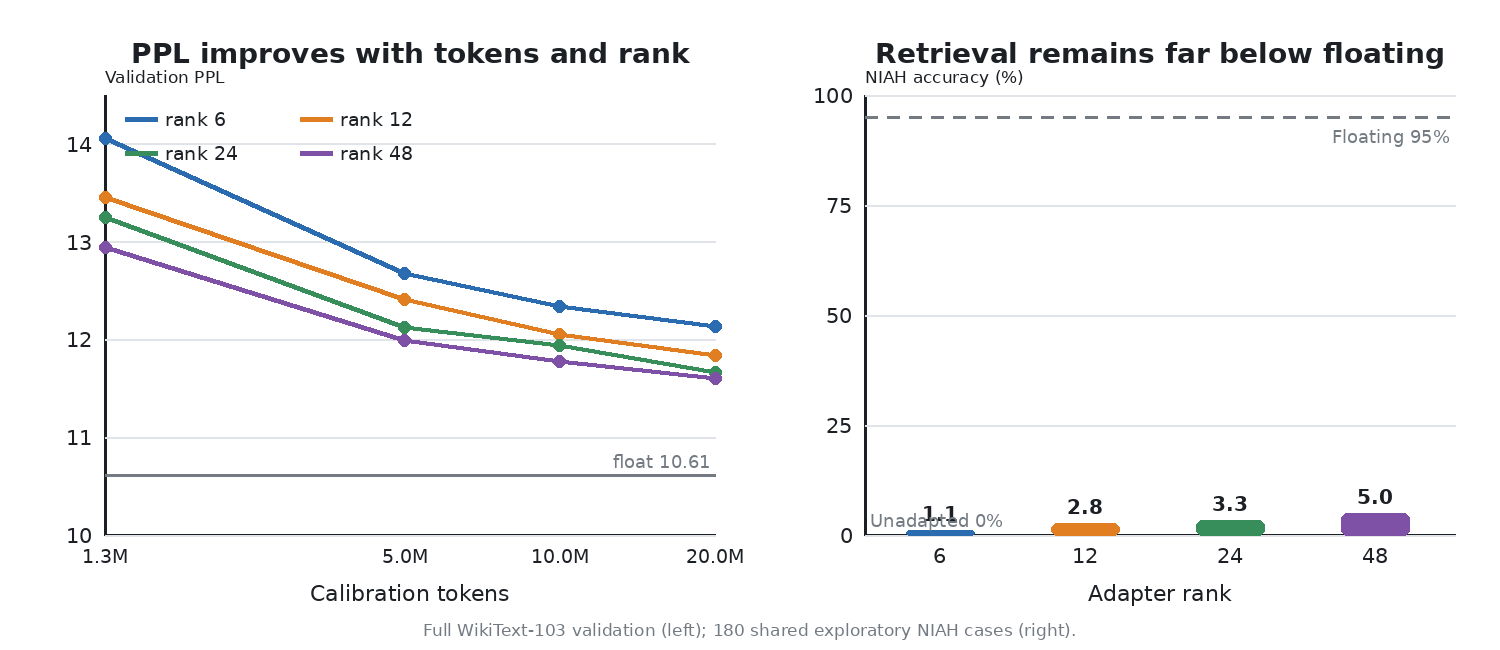}
\caption{Increasing rank and calibration monotonically improves TinyLlama's
2-bit validation PPL (left), but retrieval reaches only 1.1--5.0\% versus 95\%
with the floating cache (right). The unadapted 2-bit cache retrieves 0\%.}
\label{fig:capacity}
\end{figure}

The short 2-bit run does not identify whether training or adapter capacity is
the bottleneck. Figure~\ref{fig:capacity} separates them in a QKV grid. For
every rank, full validation PPL improves monotonically at 1.31M, 5.00M, 10.00M,
and 20.00M calibration tokens; for every token budget, increasing rank improves
PPL. Consequently every rank selects its 20.00M-token checkpoint. On the full
test split, ranks 6, 12, 24, and 48 reach PPL 11.9172, 11.6328, 11.4667, and
11.3935, respectively, compared with 576.1030 unadapted and 10.3988 floating.
For rank 48 specifically, validation PPL falls from 11.7769 at 10.00M tokens to
11.5955 at 20.00M, a further 0.1814 reduction; the tested range therefore shows
continued benefit from additional calibration rather than a clear plateau.
Additional calibration and capacity therefore bring PPL close to the floating
reference. The selected rank-48 adapter has 9.191M trainable parameters.
Repeating the 20.00M-token recipe gives test PPL $11.4000\pm0.0059$ and recovery
$99.8230\%\pm0.0010\%$ across seeds.

The same 2-bit rank ordering appears behaviorally, but at a much smaller
absolute level: the four 20M-token adapters retrieve 2, 5, 6, and 9 of 180
targets as rank increases, versus 0 unadapted and 171 floating. Ranks 24 and 48
improve over the shared unadapted cases after Holm correction across the four
one-sided exact paired tests ($p_{\mathrm{adj}}=.0469$ and $.0078$). These
exploratory results select rank 48; the three-seed confirmation below uses an
independent case set.

\subsection{Behavioral retrieval beyond perplexity}

Table~\ref{tab:niah} gives the behavioral result. Gemma's floating and packed
caches retrieve 114/120 and 112/120 targets. All three QK adapters improve the
packed count, reaching 116, 117, and 115. Paired against the same packed cases,
the adapters fix six, six, and five failures while breaking two, one, and two
successes. Each seed has few discordant cases, so these gains are descriptive.

The TinyLlama QKV suite has a larger sample and twenty cases per cell.
Its floating and packed caches retrieve 171/180 and 163/180 targets. All three adapters
improve the packed count, reaching 168, 169, and 167. Paired against the same
packed cases, they fix 10--11 failures while breaking 5--7 successes. The net
direction is consistent with perplexity recovery. No individual one-sided exact
paired test yields $p<.05$, so these gains are descriptive.

The 2-bit confirmation uses a third prompt set, unseen during rank
selection. Its floating and unadapted caches retrieve 170/180 and 0/180;
independently trained rank-48 adapters retrieve 11, 12, and 11. Every recovered
case is a fix, while there are none to break. The corresponding Holm-adjusted one-sided
exact $p$-values are $.0010$, $.0007$, and $.0010$. This confirms a nonzero,
seed-stable behavioral gain, while its 6.1--6.7\% accuracy remains far from the
94.4\% floating reference.

\begin{table}[H]
\centering
\caption{Associative NIAH first-code retrieval. Adapted columns are
independently trained adapters; cases are shared within each row.}
\label{tab:niah}
\footnotesize
\renewcommand{\arraystretch}{1.05}
\setlength{\tabcolsep}{2.0pt}
\begin{tabular*}{\columnwidth}{@{\extracolsep{\fill}}lrrrrr@{}}
\toprule
Setting & Float & Packed & Seed 0 & Seed 1 & Seed 2 \\
\midrule
Tiny QKV, 4-bit & 171/180 & 163/180 & 168/180 & 169/180 & 167/180 \\
Tiny QK, 2-bit (short) & 86/90 & 0/90 & 0/90 & -- & -- \\
Tiny QKV, 2-bit (20M) & 170/180 & 0/180 & 11/180 & 12/180 & 11/180 \\
Tiny QK, 4-bit & 86/90 & 81/90 & 86/90 & 87/90 & 81/90 \\
Tiny QK, 8-bit & 86/90 & 86/90 & 86/90 & -- & -- \\
Gemma QK, 4-bit & 114/120 & 112/120 & 116/120 & 117/120 & 115/120 \\
\bottomrule
\end{tabular*}
\end{table}

RULER provides a broader behavioral comparison (Table~\ref{tab:ruler}). The adapted Gemma cache
raises the eight-cell macro-average from 42.80 to 48.33, exceeding the 46.15
floating control on this subset. The aggregate is not uniform: multi-value
retrieval and common-word extraction improve, single-needle retrieval declines,
and variable tracking is essentially unsolved in every condition. Thus the
adapter restores useful long-context behavior, but the aggregate alone would
hide redistribution across tasks.

\begin{table}[H]
\centering
\caption{Official RULER subset scores (\%) on Gemma-4-12B. Single and Multi are
needle retrieval; VT is variable tracking and CWE common-word extraction.
Eight-cell macro means are 46.15 (floating), 42.80 (packed), and 48.33 (adapted).}
\label{tab:ruler}
\footnotesize
\renewcommand{\arraystretch}{1.05}
\begin{tabular*}{\columnwidth}{@{\extracolsep{\fill}}llrrrr@{}}
\toprule
Cache & Length & Single & Multi & VT & CWE \\
\midrule
Floating & 4K & 84 & 98 & 0.8 & 29.6 \\
         & 8K & 60 & 84 & 0.0 & 12.8 \\
\addlinespace[1pt]
Packed   & 4K & 80 & 85 & 0.8 & 37.6 \\
         & 8K & 60 & 61 & 0.0 & 18.0 \\
\addlinespace[1pt]
Adapted  & 4K & 56 & 97 & 0.0 & 69.2 \\
         & 8K & 48 & 90 & 0.0 & 26.4 \\
\bottomrule
\end{tabular*}
\end{table}

\begin{figure*}[!t]
\centering
\includegraphics[width=0.94\textwidth]{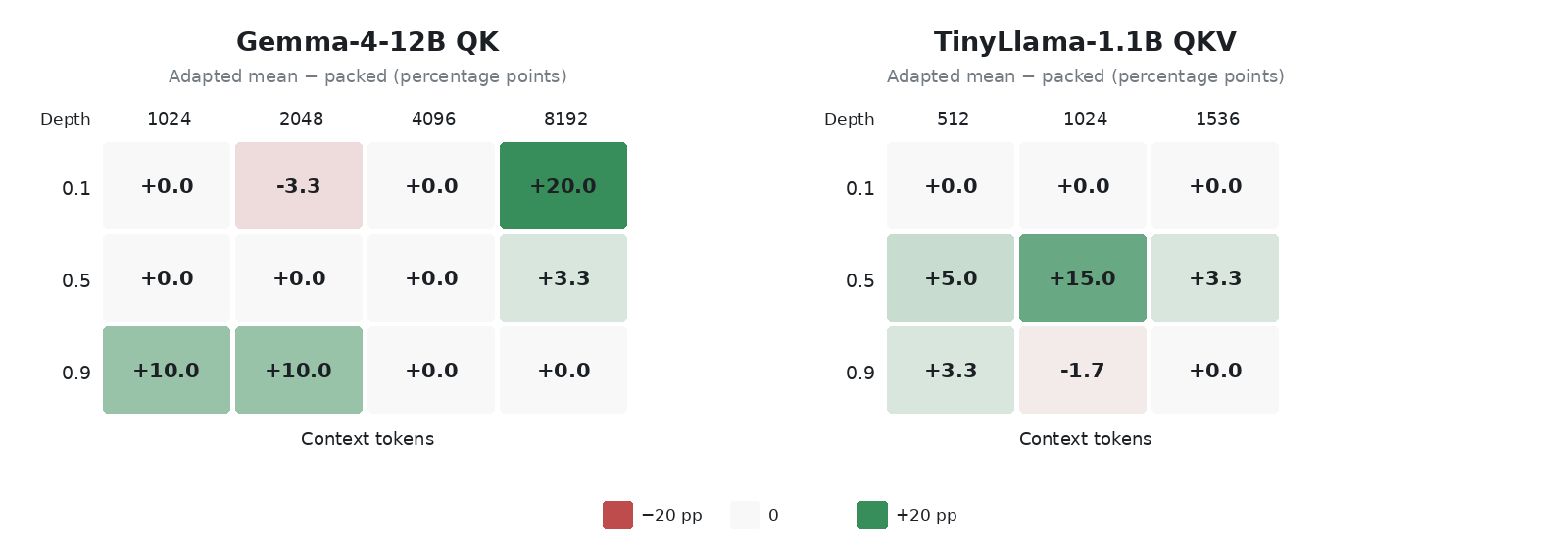}
\caption{Cell-wise change in associative retrieval from the unadapted packed
cache to the mean of three independently trained adapters. Green cells improve
and red cells decline; values give the percentage-point difference.}
\label{fig:niah}
\end{figure*}

Figure~\ref{fig:niah} shows where adaptation changes retrieval. Gemma gains 20
percentage points for shallow needles at 8,192 tokens and 10 points for deep
needles at 1,024 and 2,048 tokens, with one 3.3-point decline. TinyLlama's
largest gain is at middle depth and 1,024 tokens, where accuracy rises from
40\% to a 55\% mean. Cell patterns are descriptive because each prompt is
evaluated with three independently trained adapters.

The cross-bit comparison clarifies the 2-bit scaling result. The short QK run recovers
98.5\% of the normalized PPL gap but retrieves 0/90 targets. Scaling both QKV
capacity and calibration moves PPL much closer to floating and reproducibly
restores 11--12 targets, but leaves most retrieval behavior absent. At 8 bits,
the normalized PPL ratio is unstable while floating, packed, and adapted
retrieval all tie at 86/90. Recovery training therefore improves both metrics at
2 bits, but a large behavioral gap remains after PPL approaches floating.

\section{Discussion and Limitations}

Low-rank adaptation reproducibly recovers quality lost under a fixed packed cache.
Gemma demonstrates recovery at the 12-billion-parameter scale, while Llama-3.1
yields the same placement ordering and shows recovery with KIVI and KVarN.
RULER shows that aggregate recovery is not uniform across tasks. In the affine
2-bit stress test, increased capacity and calibration recover
11--12/180 NIAH cases, far below the 170/180 floating result.

NIAH and the four-task RULER subset measure synthetic behavior rather than
broad downstream performance. Main affine and selected TinyLlama results use three seeds;
placement, KIVI, and KVarN use one. These one-seed results do not quantify
variance or support comparisons between quantizers. Three model families do not
establish universality, Gemma has no placement sweep, and its optimizer skips a
small fraction of updates despite close seed results.

Section~3.4 separates two deployment paths: dense merging removes the auxiliary
branch, whereas the reported NF4 path retains it and adds 21.889M MACs per
processed token for Gemma. End-to-end fused-kernel benchmarks would determine
how this arithmetic maps to latency, traffic, energy, and throughput.
Compatibility with untested outlier, layer-mixed, and learned-state methods
remains theoretical, and broader downstream tasks would test how quality
recovery translates into deployment utility.

\section{Conclusion}

Across tested models, low-rank attention adapters recover quality lost to
physically packed affine, symmetric, KIVI, and KVarN caches. Matched budgets
favor QKV. Rank and calibration nearly restore 2-bit perplexity but not
retrieval; 4-bit NIAH and RULER show useful, nonuniform recovery. Together,
these results position low-rank attention adaptation as a post-training
mechanism for recovering quality under fixed cache formats.\\

All experiments and results are available at
\href{https://github.com/saifmb0/kvlora}{\texttt{github.com/saifmb0/kvlora}}.

\FloatBarrier

\end{document}